\documentclass[twocolumn,secthm,seceqn]{autart}    

\usepackage{graphicx}          

\usepackage{times} 
\usepackage{amsmath} 
\usepackage{amssymb}  

\usepackage{arydshln}

\usepackage{xcolor} 
\usepackage{mathtools}
\usepackage{algpseudocode}
\usepackage{algorithm}

\DeclarePairedDelimiter{\norm}{\lVert}{\rVert} 
\newcommand{\R}{\mathbb{R}}

\begin{document}

\begin{frontmatter}

\title{Modular Model Reduction of Interconnected Systems: A Robust Performance Analysis Perspective \thanksref{footnoteinfo}} 

\thanks[footnoteinfo]{Corresponding author: Lars A.L. Janssen.}

\author[TU/e]{Lars A.L. Janssen}\ead{l.a.l.janssen@tue.nl},    
\author[RUG]{Bart Besselink}\ead{b.besselink@rug.nl},               
\author[TU/e]{Rob H.B. Fey}\ead{r.h.b.fey@tue.nl},   
\author[TU/e]{Nathan van de Wouw}\ead{n.v.d.wouw@tue.nl}  

\address[TU/e]{Dynamics \& Control group, Department of Mechanical Engineering,  Eindhoven University of Technology}  
\address[RUG]{Bernoulli Institute for Mathematics, Computer Science and Artificial Intelligence, University of Groningen}             


\begin{abstract}                          
Many complex engineering systems consist of multiple subsystems that are developed by different teams of engineers.
To analyse, simulate and control such complex systems, accurate yet computationally efficient models are required.
Modular model reduction, in which the subsystem models are reduced individually, is a practical and an efficient method to obtain accurate reduced-order models of such complex systems.
However, when subsystems are reduced individually, without taking their interconnections into account, the effect on stability and accuracy of the resulting reduced-order interconnected system is difficult to predict.
In this work, a mathematical relation between the accuracy of reduced-order linear-time invariant subsystem models and (stability and accuracy of) resulting reduced-order interconnected linear time-invariant model is introduced.
This result can subsequently be used in two ways. 
Firstly, it can be used to translate accuracy characteristics of the reduced-order subsystem models directly to accuracy properties of the interconnected reduced-order model. 
Secondly, it can also be used to translate specifications on the interconnected system model accuracy to accuracy requirements on subsystem models that can be used for fit-for-purpose reduction of the subsystem models.
These applications of the proposed analysis framework for modular model reduction are demonstrated on an illustrative structural dynamics example.
\end{abstract}

\end{frontmatter}

\section{Introduction}
Many complex engineering systems, such as those in the automotive and high-tech industry, rely on the integration of multiple interconnected subsystems/modules. 
These subsystems are increasingly of a multiphysics and/or multidisciplinary nature and their dynamic behavior is typically developed, modelled, and analysed independently, possibly by distinct teams.
For the analysis and design of a single subsystem, detailed high-order models with a high level of complexity are typically used.
The dynamics of each subsystem may be modelled and analyzed individually in, e.g., the mechanical, electrical or thermal domain, or combinations thereof and control engineers may use these subsystem models to guarantee required dynamic behavior.
Interconnecting such high-order subsystem models to analyze the interconnected system would lead to models of such high complexity that the dynamics analysis of the interconnected system becomes infeasible.
Therefore, simplified versions of these subsystem models, i.e., reduced-order models (ROMs), are used instead for analysis of the interconnected system model~\cite{reis2008}.
In this paper, we will provide a framework for analyzing how the errors introduced by subsystem reduction influence the accuracy of the model of the interconnected system.

The general process of simplifying models, called model order reduction (MOR), is a topic that is studied in several research fields such as structural dynamics~\cite{craig2000}, systems and control~\cite{gugercin2004}, thermal systems~\cite{veldman2018}, see~\cite{besselink2013,antoulas2005,schilders2008} for overviews.
Generally, in MOR, the aim is to find a ROM that is reduced significantly in complexity while still providing an accurate description of the dynamic behaviour of the high-order model  in some frequency range of interest. 
For linear time-invariant (LTI) systems, this is performed commonly using projection-based methods~\cite{antoulas2005}.
These methods rely on the projection of the high-order model onto a subspace with a reduced number of states. 
Examples of commonly used projection-based methods used for MOR are the proper orthogonal decomposition method~\cite{kerschen2005}, reduced basis methods~\cite{boyaval2010}, balancing methods~\cite{gugercin2004,moore1981,glover1984} and Krylov methods~\cite{grimme1997}.

Applying these MOR methods on a subsystem level can lead to accurate subsystem models. 
However, it does not necessarily lead to the best approximation of the behavior of the \emph{interconnected} system.
Therefore, several other approaches have been explored.
Although direct reduction of the entire interconnected system as a whole often leads to accurate models, it completely destroys the interconnection structure~\cite{lutowska2012}.
As a solution to this problem, structure-preserving methods are available for interconnected systems~\cite{lutowska2012,sandberg2009,vandendorpe2008,schilders2014}.
In the structural dynamics field, component mode synthesis (CMS) methods are also structure-preserving \cite{klerk2008}.
However, these methods do not provide a priori error bounds.
A complementary approach is model reduction of network systems \cite{besselink2015,yeung2009,cheng2021}, where the aim is often to reduce the interconnection structure rather than the subsystem dynamics.
Furthermore, for network systems, the interconnected system typically consists of a large number of subsystems of relatively small complexity.
Therefore, these methods are not developed for the case of several highly complex interconnected LTI subsystems.

To perform accurate model reduction, knowledge on the entire interconnected system model is needed. 
However, this is typically not feasible for large-scale models of interconnected systems.
In such cases, modular model reduction is the preferred approach~\cite{vaz1990,buhr2018}, as it provides a huge computational advantage.
Additionally, it allows to decompose the overall complexity reduction problem into smaller ones tailored to the nature of the subsystems \cite{reis2008}.
Therefore, a MOR method that is best suitable for reducing a specific subsystem can be chosen for each subsystem individually \cite{benner2015}.
Furthermore, modular model reduction has the advantage that it preserves the interconnection structure and the physical interpretation of the subsystems.
Specifically, in the reduced-order model of the interconnected system, 1) the topology of the system that describes how the subsystems interact, i.e., the interconnection structure, is preserved and, 2) each subsystem model still represents the physical behaviour of that subsystem.

Unfortunately, when a system model is reduced modularly, i.e., individually on a subsystem level, it is challenging to quantify how the stability and accuracy of the reduced-order interconnected system model are affected by a loss of accuracy induced by the reduction of a subsystem model.
Although there are some a priori error bounds available in the literature~\cite{reis2008,ishizaki2013}, these are often highly conservative and therefore less suitable for competitive engineering applications.
In addition, if there are requirements on accuracy of the interconnected system it is difficult to translate these requirements to subsystem level.
Currently, to the best of the authors' knowledge, there is no method that allows to specify accuracy requirements for subsystem models from a global perspective that guarantee a given accuracy for the required overall interconnected system a priori (i.e., before performing the actual reduction).
This paper has the following contributions. The main contribution is a framework for quantitatively relating the input-to-output subsystem model accuracy to input-to-output accuracy of the interconnected system model. 
This framework is obtained by using a robust performance analysis approach in which model reduction errors are modelled as uncertainties.
It allows for a direct implementation of efficient mathematical tools from the theory of robust control~\cite{zhou1998,skogestad2007} such as the structured singular value~\cite{packard1993}.
These tools can be used to relate subsystem reduction errors to the reduction error of the interconnected system, thus allowing for analyzing and optimizing the accuracy of subsystem reduction.
As specific uses of this framework, two additional contributions follow.

First, the modular model reduction is analysed using a bottom-up approach.
This approach allows to determine the propagation of errors introduced by subsystem model reduction to the reduced-order interconnected system model.
Using this approach, a priori stability guarantees and error bounds on the interconnected system model can be computed using only (a priori) knowledge on reduction errors on a subsystem level.
In our earlier work~\cite{janssen2022}, we presented an iterative version this approach using bisection to find a priori error bounds.
In this work, these error bounds can be efficiently computed either on a frequency-dependent or global (frequency-independent) level by solving simple linear matrix inequalities (LMIs).

Second, a top-down approach is given.
In this approach, accuracy specifications on a subsystem level are determined based on requirements on the interconnected system model accuracy.
Typically, we are interested in obtaining a (reduced) model of the interconnected system that meets some specific accuracy requirements.
With this approach, by solving simple LMIs, we can translate these accuracy requirements to the subsystem level.
Therefore, model reduction can be applied on a subsystem level, which allows for a completely modular approach where the model reduction method can be specifically chosen for the individual subsystem, as long as error bound requirements are met.

The paper is organized as follows. Section~\ref{sec:Problem} gives the problem statement including the modelling framework. 
In Section~\ref{sec:Robust}, a robust performance perspective on modular model reduction is developed, i.e., it is explained how to relate subsystem model reduction errors to the reduced-order interconnected system model error and vice versa.
Specific applications of these relations, i.e., the bottom-up and top-down approaches, are given in Section~\ref{sec:Error} which are demonstrated on an illustrative structural dynamics example system in Section~\ref{sec:example}.
The conclusions and recommendations for future work are given in Section~\ref{sec:Conclusion}.

\emph{Notation}. The set of real numbers is denoted by $\mathbb{R}$. 
The set of complex numbers is denoted by $\mathbb{C}$. 
Given a vector $x \in \mathbb{C}^n$, its Euclidean norm is given as $\norm{x}$. 
Given a transfer function (matrix) $G(s)$, where $s$ is the Laplace variable, $\norm{G}_\infty$ denotes its $\mathcal{H}_\infty$-norm. 
The real rational subspace of $\mathcal{H}_{\infty}$ is denoted by $\mathcal{RH}_{\infty}$, which consists of all proper and real rational stable transfer matrices. 
Given a complex matrix $A$, $A^H$ denotes its conjugate transpose, $\bar{\sigma}(A)$ denotes its largest singular value, $\rho(A)$ denotes its spectral radius and $A = \text{diag}(A_1,A_2)$ denotes a block-diagonal matrix with submatrices $A_1$ and $A_2$. 
The identity matrix of size $n$ is denoted by $I_n$.

\section{Problem statement}
\label{sec:Problem}
\begin{figure}
   \centering
   \includegraphics[scale=1, page=6]{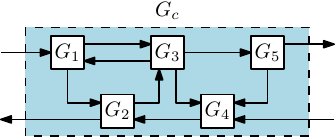} 
   \caption{Block diagram representation an arbitrarily interconnected system consisting of several subsystems.}
   \label{fig:Gc_example}
\end{figure}

In this work, we consider a set of arbitrarily interconnected LTI subsystems.
These subsystems interact by linking for each subsystem, (a part of) subsystem outputs, to (a part of) inputs of the other subsystems. 
Additionally, some subsystems will have one or multiple external input(s) and/or output(s).
An example of such a system is given in Figure~\ref{fig:Gc_example}.
This class of systems contains a wide variety of interconnected systems for which model reduction is essential to enable design, analysis and control of the system dynamics.

\subsection{Modular model framework}
\label{sec:model_framework}
Consider $k$ high-order subsystems $j \in \{1,\dots,k\}$ with (proper real rational) transfer functions $G_j(s)$, inputs $u_j$ and outputs $y_j$ of dimensions $m_j$ and $p_j$, respectively, and McMillan degree $n_j$.
We collect the subsystem transfer functions in the block-diagonal transfer function
\begin{equation}
G_b(s) := \textrm{diag}(G_1(s),\dots,G_k(s)),
\end{equation}
for which the total number of inputs and outputs are then given by $m_b := \sum_{j=1}^{k}m_j$ and $p_b := \sum_{j=1}^{k}p_j$, respectively.
We define inputs $u_b^\top := \left[u_1^\top,\dots,u_k^\top \right]$ and outputs $y_b^\top := \left[y_1^\top,\dots,y_k^\top \right]$.

\begin{figure}
   \centering
   \includegraphics[scale=1, page=1]{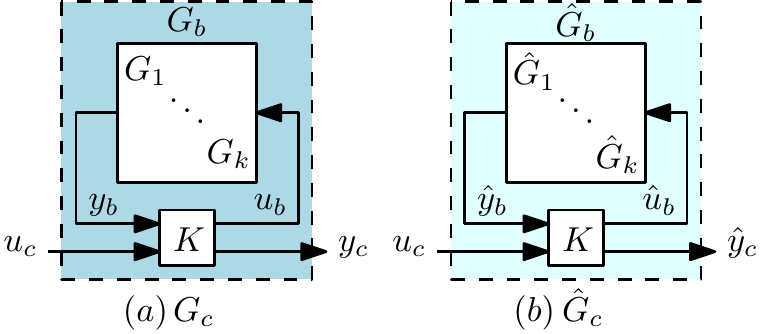} 
   \caption{Block diagram representation of (a) high-order interconnected system $G_c(s)$ and (b) reduced-order interconnected system $\hat{G}_c(s)$.  K represents a static interconnection block.}
   \label{fig:Gc}
\end{figure}

The subsystems are interconnected according to
\begin{equation}
\label{eq:connection}
\left[\begin{array}{c}
u_b \\ y_c 
\end{array}\right] = 
K \left[\begin{array}{c}
y_b \\ u_c 
\end{array}\right], \qquad K = \left[\begin{array}{cc}
K_{11} & K_{12} \\
K_{21} & K_{22}
\end{array}\right]
\end{equation}
where $u_c$ and $y_c$ denote external inputs and outputs, respectively, see Figure~\ref{fig:Gc}(a).
The number of external inputs and outputs are given by $m_c$ and $p_c$, respectively.
Then, the transfer function from $u_c$ to $y_c$ is given by the upper linear fractional transformation (LFT) of $G_b(s)$ and $K$, which yields
\begin{equation}
\label{eq:Gc}
G_c(s) := K_{21}G_b(s)( I - K_{11}G_b(s))^{-1}K_{12} + K_{22}.
\end{equation}
Throughout this paper, we make the following assumption.
\begin{assum}\label{as:mainassumption}
The system \eqref{eq:Gc} is
\begin{enumerate}
\item well-posed, i.e., $I-K_{11}G_b(s)$ has a proper real rational inverse;
\item is internally stable, i.e., $(I-K_{11}G_b(s))^{-1}\in\mathcal{RH}_{\infty}$ and $G_c(s)\in\mathcal{RH}_{\infty}$.
\end{enumerate}
\end{assum}
Note that a feedback system is defined to be well-posed if \emph{all} closed-loop transfer functions are well-defined and proper, and internally stable if \emph{all} closed-loop transfer functions are stable. 
Because $K$ is a \emph{static} interconnection matrix, the specified transfer functions in Assumption~\ref{as:mainassumption} are necessary and sufficient for their respective properties on $G_c(s)$.
For more details on well-posedness and internal stability, see~\cite[Definitions 5.1 and 5.2]{zhou1998}.
\begin{rem}
For many systems within this modelling framework, external inputs and outputs are directly connected to a subsystem input and output, respectively.
In those instances, $u_c$ will contain identical elements in $u_b$.
The same holds for output signals $y_c$ and $y_b$.
\end{rem}

\subsection{Modular model order reduction}
\label{sec:MOR}
In model order reduction, we aim to find a ROM of a system with (significantly) fewer internal states than the number of states of the high-order model.
In this paper, we compute the ROM of the system modularly, i.e., we reduce each subsystem independently. 
Therefore, we need to consider reduced-order subsystems $j \in \{1,\dots,k\}$ and their (real rational proper) transfer functions $\hat{G}_j(s)$, each with inputs $\hat{u}_j$ and outputs $\hat{y}_j$ with dimensions $m_j$ and $p_j$, respectively, and McMillan degree $r_j$.
Let the reduced-order block-diagonal transfer function be given as 
\begin{equation}
\hat{G}_b(s) := \textrm{diag}(\hat{G}_1(s),\dots,\hat{G}_k(s)). 
\end{equation}
Then, we define inputs $\hat{u}_b^\top := \left[\hat{u}_1^\top,\dots,\hat{u}_k^\top \right]$ and outputs $\hat{y}_b^\top := \left[\hat{y}_1^\top,\dots,\hat{y}_k^\top \right]$ with dimensions $m_b$ and $p_b$, respectively.
Since we only reduce the subsystem models, the interconnection structure remains preserved.
Therefore, the reduced-order interconnected system transfer function is, similar to (\ref{eq:Gc}), given by
\begin{equation}
\label{eq:hatGc}
\hat{G}_c(s) := K_{21}\hat{G}_b(s)( I - K_{11}\hat{G}_b(s))^{-1}K_{12} + K_{22}.
\end{equation}
Here, the reduced-order interconnected system has external inputs $u_c$ and external outputs $\hat{y}_c$.
The reduced-order interconnected system is illustrated in Figure~\ref{fig:Gc}(b).
Note that we do not make any assumptions on well-posedness and stability of \eqref{eq:hatGc}. In fact, we would like to find conditions on the model reduction procedure that guarantees these properties given the high-order model $G_c(s)$ satisfying Assumption~\ref{as:mainassumption}.

In this paper, the aim is to compute a ROM that can accurately describe the external input-to-output behavior of the interconnected system.
In the described modelling framework, this means that given the same external input $u_c$, the difference between the external output signal of the high-order and the reduced-order system $e_c := \hat{y}_c - y_c$, is required to be small.
Therefore, the accuracy of the reduced-order interconnected system can be described by the interconnected system error dynamics, which is defined as
\begin{equation}
\label{eq:Ec}
E_c(s) := \hat{G}_c(s) - G_c(s).
\end{equation}
Given (\ref{eq:Gc}), (\ref{eq:hatGc}) and (\ref{eq:Ec}), $E_c$ can be written as
\begin{equation}
\label{eq:Ec_ext}
\begin{aligned}
E_c(s) &= \quad K_{21}\hat{G}_b(s)( I - K_{11}\hat{G}_b(s))^{-1}K_{12}\\
 & \quad - K_{21}G_b(s)( I - K_{11}G_b(s))^{-1}K_{12}.
\end{aligned}
\end{equation}
However, with a modular approach, subsystems are reduced independently, which therefore means that knowledge on the accuracy of the reduced models is generally only available on a \emph{subsystem level}.
The accuracy of the reduced-order subsystems can be described by the subsystem error dynamics, which is defined as
\begin{equation}
\label{eq:Ej}
E_j(s) := \hat{G}_j(s) - G_j(s).
\end{equation}
The associated output error is denoted as $e_j = \hat{y}_j -y_j$. 
We assume that the reduction is such that $E_j(s)\in\mathcal{RH}_{\infty}$.
\begin{rem}
\label{rem:apriori}
For some model reduction methods, bounds on the error dynamics are available a priori. 
For example, model reduction using traditional balanced truncation\cite{moore1981,enns1984} can be applied to systems satisfying $G_j(s)\in\mathcal{RH}_{\infty}$.
It guarantees preservation of stability, i.e., $\hat{G}_j(s)\in\mathcal{RH}_{\infty}$, and therefore $E_j(s)\in\mathcal{RH}_{\infty}$.
The a priori error bound on the reduced-order subsystem $j$ is then given by
\begin{equation}
\label{eq:apriori}
\|E_j\|_\infty \leq \sum\limits_{i=r+1}^{n_j}\sigma_{j,i}.
\end{equation}
Here, $\sigma_{j,i}, i = 1,\dots,n_j$ are the Hankel singular values. 
\end{rem}
Computation of the reduction error of the interconnected system $E_c$ on the basis of knowledge of $E_j$ can be computationally expensive or even infeasible for complex interconnected high-order models. 
Moreover, this computation is only possible when the \emph{exact} subsystem reduction errors $E_j$ are known for all subsystems.

In addition, usually, requirements are posed to model accuracy and model complexity on the \emph{level of the interconnected system}.
With modular model reduction however, the subsystem models are reduced on \emph{subsystem level}.
Therefore, the need arises for establishing a relation between subsystem error dynamics $E_j$ and the interconnected system error dynamics $E_c$ without exact a priori knowledge of the specific error dynamics $E_j$ for all subsystems.
In the next section, we will show how this relation can be formulated using a robust performance analysis perspective.
Specifically, we will show how this relation gives the ability to pursue
\begin{enumerate}
\item \emph{a bottom-up approach}: evaluate the propagation of subsystem reduction errors $E_j$ to the resulting stability and accuracy of the reduced interconnected system $E_c$, and
\item \emph{a top-down approach}: determine requirements on the subsystem reduction error dynamics $E_j$ to meet requirements on stability and specified maximal error $E_c$ of the reduced interconnected system.
\end{enumerate}

\section{A robust performance perspective on modular model reduction}
\label{sec:Robust}
\begin{figure}
   \centering
   \includegraphics[scale=1, page=3]{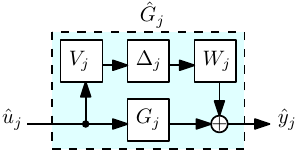} 
   \caption{Block diagram representation of $\hat{G}_j(s) = G_j(s) + E_j(s)$ where $E_j(s)$ is given as a function of $V_{j}(s)$, $W_{j}(s)$ and $\Delta_j(s)$.}
   \label{fig:hatGj}
\end{figure}

In this section, we show how reformulation of the modular model reduction framework into a robust performance analysis problem setting can lead to a directly computable relation between $E_j$ and $E_c$.
To this end, we rewrite (\ref{eq:Ej}) to 
\begin{equation}
\label{eq:hatGjDelta}
\hat{G}_j(s) = G_j(s) + E_j(s).
\end{equation}
and recall that $E_j(s)\in\mathcal{RH}_{\infty}$.
As a result, we can define weighting transfer functions $V_{j}(s)\in\mathcal{RH}_{\infty}$  and $W_{j}(s)\in\mathcal{RH}_{\infty}$ such that $E_j(s)$ can be written as
\begin{equation}
E_j(s) = W_j(s) \Delta_j(s) V_j(s),
\end{equation}
for some $\Delta_j(s) \in \mathcal{RH}_\infty$ satisfying
\begin{equation}
\|\Delta_j\|_\infty \leq 1,
\end{equation}
Then, (\ref{eq:hatGjDelta}) can be rewritten as
\begin{equation}
\label{eq:EjasDeltaj}
\hat{G}_j(s) = G_j(s) + W_{j}(s)\Delta_j(s)V_{j}(s).
\end{equation}
This representation is shown in Figure~\ref{fig:hatGj}.
Similar to before, we collect the matrices $\Delta_j$ and the weighting matrices as
\begin{equation*}
\begin{aligned}
\Delta_b(s) &:= \textrm{diag}\left(\Delta_1(s),\ldots,\Delta_k(s) \right),\\ 
V_{b}(s) &:= \textrm{diag}\left(V_{1}(s),\dots,V_{k}(s)\right), \textrm{and} \\ 
W_{b}(s) &:= \textrm{diag}\left(W_{1}(s),\dots,W_{k}(s)\right),
\end{aligned}
\end{equation*}
such that we have 
\begin{equation}
E_b(s) = \hat{G}_b(s) - G_b(s) = W_b(s)\Delta_b(s) V_b(s).
\end{equation}

\begin{figure}
   \centering
   \includegraphics[scale=1, page=2]{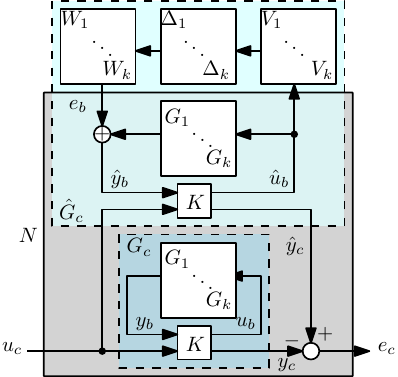} 
   \caption{Block diagram representation of the error dynamics of the interconnected system, $E_c(s) = \hat{G}_c(s) - G_c(s)$, as a function of $V_{b}(s)$, $W_{b}(s)$, $\Delta(s)$ and the nominal system $N(s)$.}
   \label{fig:Ec}
\end{figure}

By replacing $\hat{G}_b(s)$ by $G_b(s) + W_b(s) \Delta_b(s) V_b(s)$ in Figure~\ref{fig:Gc}(b) and comparing it with the high-order system $G_b(s)$ in Figure~\ref{fig:Gc}(a), we obtain the block diagram in Figure~\ref{fig:Ec}. 
This allows us to rewrite the interconnected system error dynamics $E_c(s)$ as in (\ref{eq:Ec_ext}) as an upper LFT of block-diagonal weighting transfer functions $W_b(s), V_b(s)$, the block-diagonal transfer function $\Delta_b(s)$, and the nominal transfer function $N(s)$ given by
\begin{equation}
\label{eq:N}
N(s) = \left[\begin{array}{cc}
N_{11}(s)  &  N_{12}(s) \\ 
N_{21}(s)  & O
\end{array}\right],
\end{equation}
where
\begin{equation}
\begin{aligned}
N_{11}(s) &= K_{11}(I-G_b(s)K_{11})^{-1},\\
N_{12}(s) &=(I-K_{11}G_b(s))^{-1} K_{12},\\
N_{21}(s) &= K_{21}(I-G_b(s)K_{11})^{-1}.\\
\end{aligned}
\end{equation}
Note that we have $N(s)\in\mathcal{RH}_{\infty}$ as a result of Assumption~\ref{as:mainassumption}.

\begin{figure}
   \centering
   \includegraphics[scale=1, page=4]{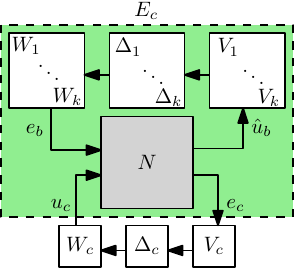} 
   \caption{Block diagram representation of the error dynamics of the interconnected system, $E_c(s) = \hat{G}_c(s) - G_c(s)$, including the nominal system $N(s)$, augmented uncertainty $\Delta_c(s)$ and weighting transfer functions $V_{c}(s)$, $W_{c}(s)$ for robust performance analysis.}
   \label{fig:Ec_perf}
\end{figure}

Figure~\ref{fig:Ec} shows the inputs and outputs of the nominal system $N(s)$, which we formulate as:
\begin{equation}
\left[
\begin{array}{c}
\hat{u}_b \\ e_c 
\end{array}
\right] = N(s)
\left[\begin{array}{c}
e_b \\ u_c 
\end{array}\right].
\end{equation}
The transfer function of the interconnected system error dynamics $E_c$ is then given by
\begin{equation}
\label{eq:EcN}
E_c(s) = N_{21} W_{b}\Delta_b V_{b}(I-N_{11}W_{b}\Delta_bV_{b})^{-1}N_{12}.
\end{equation}
As is standard within robust control theory~\cite{zhou1998}, with (\ref{eq:EcN}), in $E_c(s)$, we have now ``pulled out" the errors introduced by the reduction of subsystems $G_j(s)$ from the nominal system $N(s)$ and shifted them into $V_{b}(s)$, $W_{b}(s)$ and $\Delta_b(s)$.
The remaining system $N(s)$ consists only of high-order models $G_b(s)$ and their interconnection structure $K$, see (\ref{eq:N}), and is thus known before reduction is applied to any of the subsystems.
\begin{rem}
In robust control theory~\cite{zhou1998,skogestad2007,packard1993}, by definition, a system satisfies a robust performance criterion if for all perturbed plants within the set of uncertain system models it satisfies the given performance specifications.
In doing so, it gives a worst-case relation between local uncertainties in the system to the global performance of this system.
In this paper, by reformulation of the problem, robust performance analysis methods can be exploited to study the relation between $E_j(s)$ and $E_c(s)$.
\end{rem}
Up to now, we have extracted the errors introduced by reduction of the subsystems through the terms $\Delta_b(s)$ ,$W_b(s)$ and $V_b(s)$. 
As we aim to relate these subsystem reduction errors to the global reduction error $E_c(s)$, it will turn out to be useful to introduce a feedback between $e_c$ and $u_c$ in Figure~\ref{fig:Ec}, leading to Figure~\ref{fig:Ec_perf}, as is typical in robust performance analysis \cite[Figure~10.5]{zhou1998}.
In this setup, the relation between local and global reduction errors can be regarded as a robust performance problem.
To make this explicit, define $\Delta_c(s)$ and weighting transfer functions $V_{c}(s)$ and  $W_{c}(s)$ which represent a performance specification on the interconnected system error dynamics $E_c(s)$.
By closing this loop as in Figure~\ref{fig:Ec_perf}, the robust performance problem becomes equivalent to a robust stability problem with augmented functions $\Delta_c(s)$, $V_{c}(s)$ and $W_{c}(s)$ \cite[Theorem~10.8]{zhou1998}. 
Furthermore, we define transfer functions
\begin{equation}
\label{eq:VWDelta}
\begin{aligned}
\Delta(s) &:= \textrm{diag}\left(\Delta_b(s),\Delta_c(s)\right),\\
V(s) &:= \textrm{diag}\left(V_{b}(s),V_{c}(s)\right), \text{ and}\\
W(s) &:= \textrm{diag}\left(W_{b}(s),W_{c}(s)\right),
\end{aligned}
\end{equation}
and assume that $V, V^{-1}, W, W^{-1} \in \mathcal{RH}_\infty$, i.e., the weighting functions are bistable and biproper.

If the error dynamics $E_b(s) \in \mathcal{RH}_\infty$ and $E_c(s) \in \mathcal{RH}_\infty$ are known exactly, by definition, a solution to $\Delta_b(s)$, $V_b(s)$ and $W_b(s)$ can be found.
However, to have the ability to solve both the bottom-up and top-down problems, as described at the end of Section~\ref{sec:Problem}, we will at this point no longer assume that $E_b(s)$ and $E_c(s)$ are known (a priori).
Instead of working with given exact error dynamics $E_b(s)$ and $E_c(s)$, we consider a larger (uncertainty) set of error dynamics that contains $E_b(s)$ and $E_c(s)$.
To define this, consider the set $\mathbf{\Delta}$ as the set of complex matrices structured accordingly as
\begin{align}
\label{eq:setDelta}
\mathbf{\Delta} := \Big\{ & \text{diag}\big(\Delta_1,\dots,\Delta_k,\Delta_c \big) \ \Big| \\
& \;\; \Delta_j\in\mathbb{C}^{p_j\times m_j}, j\in\{1,\ldots,k\}, \Delta_c\in\mathbb{C}^{m_c \times p_c} \Big\}.\nonumber
\end{align}
Given $\mathbf{\Delta}$, the MOR problem is reformulated as a robust performance problem where the error dynamics $E_b(s)$ and the performance specification on $E_c(s)$ are represented as an uncertain system bounded by $\mathbf{\Delta}$ and the weighting functions $W(s)$ and $V(s)$. 
Therefore, computational tools from the field of robust control, specifically, the structured singular value $\mu$, can be used.
\begin{defn} (\cite[Definition 3.1]{packard1993}).
Given matrix $M \in \mathbb{C}^{(m_b+p_c) \times (p_b+m_c)}$, the structured singular value is
\begin{equation}
\mu_{\mathbf{\Delta}}(M) := \frac{1}{\min\left\lbrace \bar{\sigma}(\Delta) \ \middle| \ \det(I-M\Delta)=0, \Delta \in \mathbf{\Delta} \right\rbrace}.
\end{equation}
\end{defn}
Using ideas from robust performance analysis, we pose the following theorem.
\begin{thm}
\label{the:global}
Consider the system (\ref{eq:Gc}) satisfying Assumption~\ref{as:mainassumption}, biproper and bistable weighting functions (\ref{eq:VWDelta}), and the error dynamics (\ref{eq:EcN}) in Figure~\ref{fig:Ec}. The following statements are equivalent:
\begin{enumerate}
    \item For any $E_j(s)\in\mathcal{RH}_{\infty}$ satisfying 
    \begin{align}
    \|W_{j}^{-1} E_j V_{j}^{-1}\|_\infty \leq 1,
    \end{align}
    $j\in\{1,\ldots,k\}$, we have that the error dynamics (\ref{eq:EcN}) are well-posed, internally stable, and satisfy
    \begin{align}
        \|V_cE_cW_c\|_\infty \leq 1.
    \end{align}
    \item With $\mathbf{\Delta}$ as in (\ref{eq:setDelta}),
    \begin{equation}
    \label{eq:mu<1}
    \sup\limits_{\omega \in \R}\mu_{\mathbf{\Delta}}\bigl(V(i\omega)N(i\omega)W(i\omega)\bigr) < 1.
    \end{equation}    
\end{enumerate}
\end{thm}
\begin{pf}
With the following remarks, we show that the theorem becomes equivalent to the robust performance criterion, as given in~\cite[Theorem 8.7]{skogestad2007} and \cite[Theorem 10.8]{zhou1998}.
\begin{enumerate}
    \item We have that $N\in\mathcal{RH}_\infty$ as a result of Assumption~\ref{as:mainassumption} and with biproper and bistable weighting functions (\ref{eq:VWDelta}) we have $VNW\in\mathcal{RH}_\infty$.
    \item Let
\begin{align}
\mathbf{\Delta_b} := \Big\{ & \text{diag}\big(\Delta_1,\dots,\Delta_k \big) \ \Big| \nonumber \\ 
& \;\; \Delta_j\in\mathbb{C}^{p_j\times m_j}, j\in\{1,\ldots,k\} \Big\}.
\end{align}  
Then, due to the block-diagonal structure of $E_b$ and the weighting filters $V_b$ and $W_b$, we have,
\begin{align}
\label{eq:delta_b2}
    W_b^{-1}&E_bV_b^{-1}(i\omega) \in \nonumber \\ 
    & \Big\{ \Delta_b\in\mathcal{RH}_{\infty} \Big| \Delta_b(s)\in\mathbf{\Delta_b} \ \forall \ s \in \mathbb{C}\Big\}.
\end{align}
\item The system $V_cE_cW_c$ is the upper LFT of weighted nominal system $VNW$ and uncertainty $\Delta_b$ as in (\ref{eq:delta_b2}). 
\item With $\mathbf{\Delta}$ as in (\ref{eq:setDelta}), we have the augmented block structure to test the robust performance of the system in Figure~\ref{fig:Ec}. Note that Figure~\ref{fig:Ec_perf} is equivalent to \cite[Figure 10.5]{zhou1998}. 
\hfill \ \qed
\end{enumerate}
\end{pf}
Theorem~\ref{the:global} is a reformulation of the robust performance criterion using $\mu$-analysis.
Using this reformulation, Theorem~\ref{the:global} provides a worst-case relation between the $\mathcal{H}_\infty$-norm of $E_j$ for all $j \in \{1,\dots,k\}$ and $E_c$.
With weighting transfer functions $V$ and $W$ it can be computed how certain error dynamics (as described in $V_j$ and $W_j$) in subsystems $E_j$ affect the interconnected system error dynamics $E_c$ in the worst case.
Additionally, Theorem~\ref{the:global} implies that the reduced-order interconnected system $\hat{G}_c$ is stable if the high-order interconnected system $G_c$ is stable and (\ref{eq:mu<1}) is satisfied, as is shown in the following corollary.
\begin{cor}
\label{cor:stability}
Let the conditions in Theorem~\ref{the:global} holds. Then, if (\ref{eq:mu<1}) is satisfied, the reduced-order interconnected system satisfies $\hat{G}_c \in \mathcal{RH}_\infty$.
\end{cor}
\begin{pf}
We have that under the conditions in Theorem~\ref{the:global}, 
\begin{enumerate}
    \item $G_c \in \mathcal{RH}_\infty$, and
    \item $E_c \in \mathcal{RH}_\infty$ if (\ref{eq:mu<1}) is satisfied.
\end{enumerate} 
Therefore, we have that the parallel connection $\hat{G}_c = G_c + E_c \in \mathcal{RH}_\infty$.
\hfill \ \qed \end{pf}
Furthermore, rephrasing Theorem~\ref{the:global} such that it provides similar bounds on a frequency-dependent level can now be formalized in the following theorem.
\begin{thm}
\label{the:local}
Consider the system (\ref{eq:Gc}) satisfying Assumption~\ref{as:mainassumption}, biproper and bistable weighting functions (\ref{eq:VWDelta}), and the error dynamics (\ref{eq:EcN}) in Figure~\ref{fig:Ec}. Let $\omega\in\R$.
Then, the following statements are equivalent:
\begin{enumerate}
    \item For any $E_j(i\omega)$ satisfying 
    \begin{align}
    \bar{\sigma}\bigl(W_{j}^{-1}(i\omega) E_j(i\omega) V_{j}^{-1}(i\omega)\bigr) \leq 1, 
    \end{align}
    $j\in\{1,\ldots,k\}$, we have that the error dynamics (\ref{eq:EcN}) satisfy
    \begin{align}
    \bar{\sigma}\bigl(V_{c}(i\omega)  E_c(i\omega) W_{c}(i\omega)\bigr) < 1
    \end{align}
    \item With $\mathbf{\Delta}$ as in (\ref{eq:setDelta}),
    \begin{equation}
    \label{eq:mu<1local}
    \mu_{\mathbf{\Delta}}\bigl(V(i\omega)N(i\omega)W(i\omega)\bigr) < 1.
    \end{equation}    
\end{enumerate}
\end{thm}
\begin{pf}
This follows similarly to the proof of Theorem~\ref{the:global}. 
However, here, \cite[Theorem 10.8]{zhou1998} is applied for each frequency individually, such as for example in \cite[Example 10.4]{zhou1998}.
\hfill \ \qed \end{pf}
Note that for Theorem~\ref{the:local}, we lose the guarantees on well-posedness and internal stability as $\mu_{\mathbf{\Delta}}$ is only computed for each frequency individually.
To guarantee well-posedness and internal stability of the system, satisfying (\ref{eq:mu<1}) of Theorem~\ref{the:global} is sufficient.

Both Theorem~\ref{the:global} and~\ref{the:local} characterize a flexible relation between $E_j$ and $E_c$.
However, computing $\mu_{\mathbf{\Delta}}$ has been established to be a NP-hard problem~\cite{young1991}. 
Fortunately, computing an upper bound on $\mu_{\mathbf{\Delta}}$ is possible and sufficient to satisfy the condition on $\mu_{\mathbf{\Delta}}$ in Theorems~\ref{the:global} and \ref{the:local}~\cite{skogestad2007}.

To this end, consider the set of scaling matrices given as
\begin{equation}
\label{eq:D}
\begin{aligned}
\mathbf{D} = \Big\{ (D_\ell, D_r) \ \Big| & D_\ell = \textrm{diag}\left( d_1 I_{p_1},\dots,d_{k} I_{p_k}, d_cI_{m_c} \right), \\
& D_r := \textrm{diag}\left( d_1 I_{m_1},\dots,d_{k} I_{m_k}, d_cI_{p_c} \right), \\
&  d_1,\dots,d_k, d_c \in \R_{>0} \Big\}. 
\end{aligned}
\end{equation}
We now formulate the following theorem, which is a slight extension of~\cite[Theorem 3.9]{packard1993}.
\begin{thm}
\label{the:LMI}
Let $M \in \mathbb{C}^{(m_b+p_c) \times (p_b+m_c)}$. 
If there exists a $(D_\ell, D_r) \in \mathbf{D}$ such that
\begin{equation}
\label{eq:UBLMI}
M  D_rM^H \prec D_\ell,
\end{equation}
then, given $\mathbf{\Delta}$ as in (\ref{eq:setDelta}),
\begin{equation}
\mu_{\mathbf{\Delta}}(M) < 1.
\end{equation}
\end{thm}
\begin{pf}
For any $(D_\ell, D_r) \in \mathbf{D}$, we have, as given in~\cite{packard1993}, the upper bound
\begin{equation}
\mu_{\mathbf{\Delta}}(M) \leq \bar{\sigma} (D_\ell^{-\frac{1}{2}}MD_r^{\frac{1}{2}}).
\end{equation}
Therefore, to verify $\mu_{\mathbf{\Delta}}(M) < 1$ it is sufficient to find some $( D_\ell, D_r ) \in \mathbf{D}$ for which $\bar{\sigma} (D_\ell^{-\frac{1}{2}}MD_r^{\frac{1}{2}}) < 1$.
Note that this is equivalent to the LMI
\begin{equation}
(D_\ell^{-\frac{1}{2}}MD_r^{\frac{1}{2}})(D_\ell^{-\frac{1}{2}}MD_r^{\frac{1}{2}})^H \prec I
\end{equation}
and therefore to the LMI
\begin{equation}
\label{eq:LMI2}
D_\ell^{-\frac{1}{2}}MD_rM^H D_\ell^{-\frac{1}{2}} \prec I.
\end{equation}
Finally, we can left- and right-multiply both sides of (\ref{eq:LMI2}) with $D_\ell^{\frac{1}{2}}$ to find (\ref{eq:UBLMI}) in the statement of the theorem.
\hfill \ \qed \end{pf}
Note that $V(i\omega)N(i\omega)W(i\omega) \in \mathbb{C}^{(m_b+p_c) \times (p_b+m_c)}$ and let $\omega \in \R$.
It follows directly from Theorem~\ref{the:LMI} that if there exists a $(D_\ell, D_r) \in \mathbf{D}$ for which
\begin{equation}
\label{eq:LMI_N}
V(i\omega)N(i\omega)W(i\omega) D_rW^H(i\omega)N^H(i\omega)V^H(i\omega) \prec D_\ell,
\end{equation}
then
\begin{equation}
\mu_{\mathbf{\Delta}}(V(i\omega)N(i\omega)W(i\omega)) < 1.
\end{equation}
For any fixed $V$, $W$ and $N$, verifying (\ref{eq:LMI_N}) can be done using standard LMI solvers.
\begin{rem}
Note that condition (\ref{eq:mu<1}) in Theorem~\ref{the:global} requires taking a supremum over $\omega$.
Several approaches are available to avoid the computational cost of computing $\mu_{\mathbf{\Delta}}\left(V(i\omega)N(i\omega)W(i\omega)\right)$ or its upper bound for all $\omega \in \R$.
Some of these methods are discussed in detail in \cite[Section 10]{packard1993}.
\end{rem}
Bringing specific structure to $V$ and $W$, the LMI in (\ref{eq:LMI_N}) can be solved such that relations between $E_j$ for all $j \in \{1,\dots,k\}$ and $E_c$ can be directly computed.
These relations provide for instance an error bound on the reduced-order interconnected system model given error bounds of the reduced-order subsystem models (bottom-up approach), and may be used to improve the reduced-order interconnected system model accuracy by reducing subsystem models to satisfy accuracy requirements on the reduced-order interconnected system model (top-down approach). 
In the next section, guidelines are given on how $V$ and $W$ can be designed specifically for these purposes in the scope of modular model reduction.

\section{Error analysis for modular model reduction}
\label{sec:Error}
With the relations given in Theorems~\ref{the:global} and \ref{the:local}, the weighting transfer functions $V$ and $W$ can be used to compute how subsystem error dynamics $E_j$ and the interconnected system error dynamics $E_c$ are related to each other, by checking if (\ref{eq:mu<1}) and (\ref{eq:mu<1local}) hold for Theorem~\ref{the:global} and \ref{the:local}, respectively.
In this section, we show several approaches to analyse how subsystem error bounds relate to the interconnected system error bounds.
These approaches rely on obtaining these relations by imposing a specific structure on the weighting transfer functions $V$ and $W$.
It is shown how relations between bounds on $E_j$ and $E_c$ can be found on a global and frequency-dependent level.
These relations will be used for the bottom-up and the top-down approaches, as indicated at the end of Section~\ref{sec:Problem}.

\subsection{Bottom-up approach: Error bounds on $E_c$ given error bounds on $E_j$}
\label{sec:bottomup}
In this section, we show how to find global and frequency-dependent a priori bounds on the interconnected error dynamics $E_c$ given bounds on the error dynamics $E_j$ for all $j \in \{1,\dots,k\}$ introduced by reduction of subsystems, and based on Theorems~\ref{the:global}, \ref{the:local} and \ref{the:LMI}.
In the bottom-up approach, we adopt the assumption that for each subsystem $j \in \{1,\dots,k\}$ we have either an a priori global (frequency-independent) error bound or a frequency-dependent error bound.
We then aim to find in the former case a global error bound and in the latter case a frequency-dependent error bound on the interconnected system by proper choices of weighting functions.

The following theorem can be used to compute a global bound $\bar{\varepsilon}_c$ on the interconnected system reduction error, i.e., we have that $\|E_c\|_\infty \leq \bar{\varepsilon}_c$ given $\|E_j\|_\infty \leq \bar{\varepsilon}_j$ for all $j \in \{1,\dots,k\}$.
\begin{thm}
\label{the:BU_GLOBAL}
Let $E_j \in \mathcal{RH}_\infty$ and let $\bar{\varepsilon}_j$ be such that $\|E_j\|_\infty \leq \bar{\varepsilon}_j$ for all $j \in \{1,\dots,k\}$. 
Consider the optimization problem
\begin{equation}
\label{eq:sdp_bottom_up}
\begin{array}{rl}
\textrm{given} \quad & \bar{\varepsilon}_{j}\ \forall \ j \in \{1,\dots,k\}\\
\textrm{minimize} \quad & \bar{\varepsilon}_c\\
\textrm{subject to} \quad & N (i\omega) W^G D_rW^G N^H(i\omega) \prec D_\ell \ \forall \ \omega,\\
& (D_\ell, D_r) \in \mathbf{D}
\end{array}
\end{equation}
with $W^G := \text{diag}(\bar{\varepsilon}_1 I_{m_1},\dots,\bar{\varepsilon}_kI_{m_k}, \bar{\varepsilon}_c^{-1}I_{p_c})$ and $D$ as in (\ref{eq:D}).
If $\bar{\varepsilon}_c^\star$ is a feasible solution to~(\ref{eq:sdp_bottom_up}), then
\begin{enumerate}
\item $E_c$ is well-posed,
\item $E_c$ is internally stable, and
\item $\|E_c\|_\infty < \bar{\varepsilon}_c^\star$.
\end{enumerate}
\end{thm}
\begin{pf}
The proof follows from Theorems~\ref{the:global} and \ref{the:LMI}. 
By substitution of $V = I_{p_b+m_c}$ and $W = W^G$ for the weighting transfer functions in Theorem~\ref{the:global}, we have that the feedback system as shown in Figure~\ref{fig:Ec_perf} is well-posed, internally stable, and
\begin{equation}
\label{eq:BU}
\begin{aligned} 
\|E_c\|_\infty &< \bar{\varepsilon}_c \text{ for all}\\
\|E_j\|_\infty &\leq \bar{\varepsilon}_j, \ \forall \ j \in \{1,\dots,k\}, 
\end{aligned}
\end{equation}
if and only if 
\begin{equation}
\label{eq:muWg}
\sup\limits_{\omega \in \R}\mu_{\mathbf{\Delta}}(N(i\omega)W^G) < 1.
\end{equation}
Additionally, it follows from Theorem~\ref{the:LMI} that (\ref{eq:muWg}) is satisfied if, for all $\omega \in \R$, there exists a $(D_\ell, D_r) \in \mathbf{D}$ such that
\begin{equation}
N (i\omega) W^G D_rW^G N^H(i\omega) \prec D_\ell
\end{equation}
which is guaranteed by the constraint in (\ref{eq:sdp_bottom_up}). This completes the proof.
\hfill \ \qed \end{pf}
Theorem~\ref{the:BU_GLOBAL} provides a method to guarantee the stability of the reduced-order interconnected system and compute a global upper bound to the $\mathcal{H}_\infty$-norm of the error dynamics of the interconnected system $\|E_c\|_\infty \leq \bar{\varepsilon}_c^\star$ introduced by reduction errors (globally) bounded by $\|E_j\|_\infty \leq \bar{\varepsilon}_j$ for all $j \in \{1,\dots,k\}$. 
\begin{rem}
\label{rem:d_select}
If we multiply the scaling matrices $(D_\ell, D_r) \in \mathbf{D}$ by any scalar $\alpha > 0$, the resulting scaled largest singular value for any matrix $M$ is given by
\begin{equation}
\bar{\sigma}\left(\frac{1}{\sqrt{\alpha}} D^{-\frac{1}{2}}_\ell M\sqrt{\alpha} D^{\frac{1}{2}}_r\right) = \bar{\sigma}\left(D^{-\frac{1}{2}}_\ell M D^{\frac{1}{2}}_r\right).
\end{equation}
Therefore, if we set any single $d_j$ or $d_c$ in (\ref{eq:D}) to a fixed value greater than zero, which can be achieved by choosing an appropriate value for $\alpha$, the upper bound to the structured-singular value remains unchanged.
\end{rem}
To solve (\ref{eq:sdp_bottom_up}), we select $d_c = 1$ as this does not change the upper bound on $\mu_\mathbf{\Delta}$ (see Remark~\ref{rem:d_select}).
In this case, we have
\begin{equation}
\label{eq:linear_part}
W^G D_r W^G = \text{diag}\left(d_1\bar{\varepsilon}_1^2I_{m_1},\dots,d_k\bar{\varepsilon}_k^2I_{m_2},\frac{1}{\bar{\varepsilon}_c^{2}} I_{p_c}\right).
\end{equation}
As the decision variable of the optimization problem~(\ref{eq:sdp_bottom_up}) appears linearly in~(\ref{eq:linear_part}) after setting $\gamma := \bar{\varepsilon}_c^{-2}$, the constraint in~(\ref{eq:sdp_bottom_up}) is a linear matrix inequality. 
The solution to problem (\ref{eq:sdp_bottom_up}) can then be computed directly using standard semidefinite programming (SDP) solvers by maximizing over $\gamma$.
Note that if no feasible solution can be found, neither well-posedness, stability, nor an error bound can be guaranteed.

For a frequency-dependent relation between $E_j$ and $E_c$, we introduce the following theorem.
\begin{thm}
\label{the:BU_FREQ}
Let $\omega\in\R$. Let $\varepsilon_j(\omega)$ be such that $\bar{\sigma}(E_j(i\omega)) \leq \varepsilon_j(\omega)$ for all $j \in \{1,\dots,k\}$.
Consider the optimization problem
\begin{equation}
\label{eq:sdp_bottom_up_freq}
\begin{array}{rl}
\textrm{given} \quad & \varepsilon_{j}(\omega) \ \forall \ j \in \{1,\dots,k\}\\
\textrm{minimize} \quad & \varepsilon_c(\omega)\\
\textrm{subject to} \quad & N (i\omega) W^F(\omega) D_rW^F(\omega) N^H(i\omega) \prec D_\ell, \\
& (D_\ell, D_r) \in \mathbf{D}
\end{array}
\end{equation}
with $W^F(\omega) := \text{diag}(\varepsilon_1(\omega) I_{m_1},\dots,\varepsilon_k(\omega)I_{m_k}, \varepsilon_c^{-1}(\omega)I_{p_c})$ and $D$ as in (\ref{eq:D}). 
If $\varepsilon_c^\star(\omega)$ is a feasible solution to~(\ref{eq:sdp_bottom_up_freq}), then $\bar{\sigma}(E_c(i\omega)) < \varepsilon_c(\omega)$.
\end{thm}
\begin{pf}
The proof follows from Theorems~\ref{the:local} and \ref{the:LMI}. 
By substitution of $V = I_{p_b+m_c}$ and $W(i\omega) = W^F(\omega)$ in Theorem~\ref{the:local}, we obtain
\begin{equation}
\label{eq:freqrel1}
\begin{aligned}
\bar{\sigma}\left(E_c(i\omega)\right) &< \varepsilon_c(\omega) \text{ for all}\\
\bar{\sigma}\left(E_j(i\omega)\right) &\leq \varepsilon_j(\omega)\ \forall \ j \in \{1,\dots,k\},
\end{aligned}
\end{equation}
if and only if 
\begin{equation}
\mu_{\mathbf{\Delta}}\bigl(N(i\omega)W^F(\omega)\bigr) < 1.
\end{equation} 
Furthermore, it follows from Theorem~\ref{the:LMI} that 
\begin{equation}
\mu_{\mathbf{\Delta}}(N(i\omega)W^F(\omega)) < 1
\end{equation}
if there exists a $(D_\ell, D_r) \in \mathbf{D}$ such that
\begin{equation}
N (i\omega) W^Fi\omega) D_r W^F(\omega) N^H(i\omega) \prec D_\ell
\end{equation}
which is guaranteed by the constraint in (\ref{eq:sdp_bottom_up_freq}). This completes the proof.
\hfill \ \qed \end{pf}
Theorem~\ref{the:BU_FREQ} provides an upper bound to the largest singular value of the error dynamics of the interconnected system $\bar{\sigma}(E_c(i\omega)) \leq \varepsilon_c^\star(\omega)$ introduced by reduction errors bounded by $\bar{\sigma}(E_j(i\omega)) \leq \varepsilon_j(\omega)$ for all $j \in \{1,\dots,k\}$ at frequency $\omega$.
Similar to the global case, we can solve the problem (\ref{eq:sdp_bottom_up_freq}) after several steps explained next.
First, by setting $d_c = 1$, we can replace $W^F(i\omega) D_r W^F(i\omega)$ by
\begin{equation}
\label{eq:linear_part_freq}
\text{diag}\left(d_1\varepsilon_1^2(\omega)I_{m_1},\dots,d_k\varepsilon_k^2(\omega)I_{m_2},\frac{1}{\varepsilon_c^{2}(\omega)} I_{p_c}\right).
\end{equation}
Then, after defining $\gamma := \varepsilon_c^{-2}(\omega)$, the inequalities in (\ref{eq:sdp_bottom_up_freq}) are linear in $\gamma$ and the solution to problem (\ref{eq:sdp_bottom_up_freq}) can be found using standard SDP solvers for any $\omega \in \R$.
Note, also for this case, it holds that if no feasible solution can be found for some $\omega \in \R$, no conclusion on the existence of an upper bound on $\bar{\sigma}(E_c(i\omega))$ for that frequency can be made.
\begin{rem}
\label{rem:bt_fd}
For the reduction of subsystems, model reduction methods such as balanced truncation can only provide global error bounds $\bar{\varepsilon}_{j}$ as shown by (\ref{eq:apriori}) in Remark~\ref{rem:apriori}.
In this case, let $\varepsilon_{j}(\omega) = \bar{\varepsilon}_{j}$ for all $j \in \{1,\dots,k\}$ for all $\omega \in \R$ using the given a priori error bounds on the subsystems $\bar{\varepsilon}_{j}$.
However, $\mu$-analysis is inherently frequency-dependent.
Therefore, a frequency-dependent error bound on the interconnected system error dynamics $\varepsilon_c(\omega)$ can still be found by using Theorem~\ref{the:BU_FREQ}.
Note that, as we will show in Section~\ref{sec:example}, the conservativeness of $\varepsilon_c(\omega)$ is subject to the conservativeness the subsystem error bound $\bar{\varepsilon}_{j}$ provided.
\end{rem}

\subsection{Top-down approach: $E_j$ specification based on $E_c$ requirements}
\label{sec:topdown}
In this section, we show how Theorems~\ref{the:global} and \ref{the:local} can be employed such that global and frequency-dependent reduction error specifications on $E_j$ on a subsystem level can be directly computed from a reduction error $E_c$ requirement on the interconnected system model.
This approach allows for specifically tailored subsystem reduction that guarantees the required accuracy on the interconnected system model.
\begin{rem} 
\label{rem:singless}
In the top-down approach, we assume that we have some reduction error bound requirement on the interconnected system, either a global bound $\bar{\varepsilon}_c$ or a frequency-dependent bound $\varepsilon_c(\omega)$.
This bound can be defined from requirements on the accuracy of the interconnected system model.
The goal in the general top-down approach is to find some set of either global $\bar{\varepsilon}_j$ or frequency-dependent $\varepsilon_j(\omega)$ reduction error bounds for all subsystems $j \in \{1,\dots,k\}$ for which it can be guaranteed that reduction the error bound on the interconnected system will not be exceeded.
In practice, there is an infinite number of possible combinations of $\bar{\varepsilon}_j$ (or $\varepsilon_j(\omega)$) that satisfy this requirement.
Given the fact that there are many subsystem reduction options to achieve the required subsystem accuracy, finding some, in some sense, to be defined, (sub-)optimal distributions of $\bar{\varepsilon}_j$ and $\varepsilon_j(\omega)$ for now requires a heuristic approach.
The development of a systematic approach to tackle this is still an open problem.
\end{rem} 
Within the scope of this paper, we assume that a global error specification $\bar{\varepsilon}_c$ on the reduction error dynamics of the interconnected system model is given.
We then focus on restricting how $E_q$, the error of a single subsystem $j=q$ contributes to the error of the interconnected system $E_c$.
This approach can be used for all subsystems individually, and therefore can be used to reduce all subsystems, but does not solve the problem of finding an optimal distribution between subsystem errors (see Remark~\ref{rem:singless}). 
However, it is still a relevant problem, since it allows for the specification of reduction error bounds on a subsystem level, based on specifications on the desired accuracy of the overall interconnected system model. 

Let reduction error bounds be given for all other reduced-order subsystems, i.e., $\|E_{j}\|_\infty \leq \bar{\varepsilon}_{j}, j\neq q$. 
Then, we aim to find the maximum to a global error bound $\bar{\varepsilon}_q$ such that $E_c$ is guaranteed to be stable and the global error specification $\bar{\varepsilon}_c$ for the interconnected system is satisfied using the following theorem.
\begin{thm}
\label{the:TD_GLOBAL}
Let $\bar{\varepsilon}_c > 0$ be given. Let $E_{j}\in \mathcal{RH}_\infty$ and let $\bar{\varepsilon}_{j}$ such that $\|E_{j}\|_\infty \leq \bar{\varepsilon}_{j}$ for all $j\neq q$. 
Consider the optimization problem
\begin{equation}
\label{eq:sdp_top_down_single}
\begin{array}{rl}
\textrm{given} \quad & \bar{\varepsilon}_c, \bar{\varepsilon}_{j}, j\neq q\\
\textrm{maximize} \quad & \bar{\varepsilon}_q\\
\textrm{subject to} \quad & N(i\omega) W^G D_rW^G N^H(i\omega) \prec D_\ell \ \forall \ \omega,\\
& (D_\ell, D_r) \in \mathbf{D}
\end{array}
\end{equation}
with $W^G := \text{diag}(\bar{\varepsilon}_1 I_{m_1},\dots,\bar{\varepsilon}_kI_{m_k}, \bar{\varepsilon}_c^{-1}I_{p_c})$ and $D$ as in (\ref{eq:D}).
If $\bar{\varepsilon}_q^\star$ is a feasible solution to (\ref{eq:sdp_top_down_single}), then for all $E_q \in \mathcal{RH}_\infty$ such that $\|E_q\|_\infty \leq \bar{\varepsilon}_q^\star$, we have 
\begin{enumerate}
\item $E_c$ is well-posed,
\item $E_c$ is internally stable, and
\item $\|E_c\|_\infty < \bar{\varepsilon}_c^\star$.
\end{enumerate}
\end{thm}
\begin{pf}
The proof follows directly from the proof of Theorem~\ref{the:BU_GLOBAL}. 
\hfill \ \qed \end{pf}
Theorem~\ref{the:TD_GLOBAL} provides a guarantee that if $\bar{\varepsilon}_q^\star$ exists, all error dynamics $E_q$ introduced by reduction of subsystem $q$ satisfying $\|E_q\|_\infty \leq \bar{\varepsilon}_q^\star$ result in stable interconnected system model reduction error dynamics bounded by $\|E_c\|_\infty < \bar{\varepsilon}_c$.
The problem (\ref{eq:sdp_top_down_single}) can be simplified similar to the global bottom-up problem as in (\ref{eq:sdp_bottom_up}).
Here, we set $d_q = 1$ and set $\gamma := \bar{\varepsilon}_q^{2}$.
Then, the inequality in (\ref{eq:sdp_top_down_single}) is linear in $\gamma$ and maximizing $\gamma$ using SDP gives a maximum global upper bound on the (allowed) subsystem error dynamics $\|E_q\|_\infty \leq \bar{\varepsilon}_q$.
If some reduced-order subsystem $\hat{G}_q$ is found for which the upper bound $\bar{\varepsilon}_q$ is satisfied, Theorem~\ref{the:TD_GLOBAL} guarantees that $\|E_c\|_\infty \leq \bar{\varepsilon}_c$.

This top-down problem is easily translated to a frequency-dependent problem using Theorem~\ref{the:local}.
Namely, we consider some frequency-dependent error specification $\varepsilon_c(\omega)$ for which we can guarantee that $\bar{\sigma}(E_c(i\omega)) \leq \varepsilon_c(\omega)$.
Additionally, we assume that some subsystem error bound $\bar{\sigma}(E_{j}(i\omega)) \leq \varepsilon_{j}(\omega)$ of all subsystems $j\neq q$ is known.
Then, we aim to find a frequency-dependent error specification $\varepsilon_q(\omega)$ for which it holds that $\bar{\sigma}(E_q(i\omega)) \leq \varepsilon_q(\omega)$ using the following theorem.
\begin{thm}
\label{the:TD_FREQ}
Let $\omega\in\R$. Let $\varepsilon_{j}(\omega)$ be such that $\bar{\sigma}(E_{j}(i\omega)) \leq \varepsilon_{j}(\omega)$ for all $j\neq q$.
Consider the optimization problem
\begin{equation}
\label{eq:sdp_top_down_single_f}
\begin{array}{rl}
\textrm{given} \quad & \varepsilon_c(\omega), \varepsilon_{j}(\omega), j\neq q\\
\textrm{maximize} \quad & \varepsilon_q(\omega)\\
\textrm{subject to} \quad & N(i\omega) W^F(\omega) D_rW^F(\omega) N^H(i\omega) \prec D_\ell,\\
& (D_\ell, D_r) \in \mathbf{D}
\end{array}
\end{equation}
with $W^F(\omega) := \text{diag}\bigl(\varepsilon_1(\omega) I_{m_1},\dots,\varepsilon_k(\omega)I_{m_k}, \varepsilon_c^{-1}(\omega)I_{p_c}\bigr)$ and $D$ as in (\ref{eq:D}).
If $\varepsilon_q^\star(\omega)$ is a feasible solution to (\ref{eq:sdp_top_down_single_f}), then for all $E_q$ such that $\bar{\sigma}(E_q(i\omega)) \leq \varepsilon_q^\star(\omega)$, we have $\bar{\sigma}(E_c(i\omega)) < \varepsilon_c(\omega)$.
\end{thm}
\begin{pf}
The proof follows directly from the proof of Theorem~\ref{the:BU_FREQ}. 
\hfill \ \qed \end{pf}
Theorem~\ref{the:TD_FREQ} provides a guarantee that if $\varepsilon_q^\star(\omega)$ exists for $\omega$, then any error dynamics at $E_q(i\omega)$ introduced by reduction of subsystem $q$ satisfying $\bar{\sigma}(E_q(i\omega)) \leq \varepsilon_q^\star(\omega)$ results in interconnected system error dynamics bounded by $\bar{\sigma}(E_c(i\omega)) < \varepsilon_c(\omega)$ at frequency $\omega$.

The problem (\ref{eq:sdp_top_down_single_f}) can be simplified similar to the frequency-dependent bottom-up problem as in (\ref{eq:sdp_bottom_up_freq}).
Here, we set $d_q = 1$ and set $\gamma := \varepsilon_q^{2}(\omega)$.
Then, problem (\ref{eq:sdp_top_down_single_f}) is linear in $\gamma$ and maximizing $\gamma$ using SDP for any $\omega \in \R$ gives a frequency-dependent upper bound on the subsystem error dynamics $\bar{\sigma}(E_q(i\omega)) \leq \varepsilon_q(\omega)$.
If some reduced-order subsystem $\hat{G}_q$ is found for which this upper bound is satisfied, Theorem~\ref{the:TD_FREQ} guarantees that $\bar{\sigma}(E_c(i\omega)) \leq \varepsilon_c(\omega)$. 
By computing (\ref{eq:sdp_bottom_up_freq}) over a frequency grid, this guarantee holds for the $\omega \in \R$ of interest.
\begin{rem}
\label{rem:e_select}
For the top-down approach, a frequency-dependent error bound $\bar{\sigma}(E_c(i\omega)) \leq \varepsilon_c(\omega)$ can be defined by the user based on requirements on the interconnected model accuracy.
This allows for the flexibility to design specifications on the interconnected system such that the reduced-order system is accurate around frequencies that are relevant for the way the model is used.
As an example, if the interconnected system model needs to be especially accurate around a certain frequency, $\varepsilon_c(\omega)$ can be defined such that it is low around this frequency.
After applying the top-down approach, meeting the specification of frequency dependent error bounds on subsystem level $\bar{\sigma}(E_q(i\omega)) \leq \varepsilon_q(\omega)$ guarantees accuracy around this frequency.
\end{rem}

In this section, we have given several approaches to compute a relation between bounds on $E_j$ and $E_c$ using $\mu$-analysis.
To properly illustrate how these approaches can be useful for modular model reduction of systems of interconnected LTI systems, in the next section, an illustrative example from structural dynamics on which these approaches are applied will be discussed.

\section{Illustrative example}
\label{sec:example}
\begin{figure}
   \centering
   \includegraphics[scale=1, page=5]{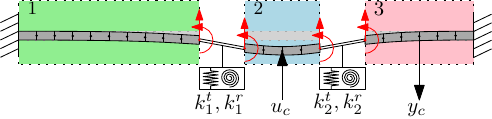} 
   \caption{Example system: two cantilever beams (Subsystems 1 and 3) connected on their free ends to a free-free beam (Subsystem 2) with translational and rotational springs.}
   \label{fig:example}
\end{figure}

To illustrate the proposed framework for error analysis of modular model reduction of interconnected systems, we apply it to a mechanical system consisting of three interconnected beams as illustrated schematically in Figure~\ref{fig:example}.
Subsystems 1 and 3 are cantilever beams which are connected on their free ends to free-free beam 2 with translational and rotational springs.
The stiffness of both translational interconnecting springs is $k_1^t = k_2^t = 4 \times 10^4$ N/m.
The stiffness of both rotational interconnecting springs is $k_1^r = k_2^r = 4 \times 10^2$ Nm/rad.
The interconnection structure matrix $K$ in (\ref{eq:connection}) is therefore given by 
\begin{equation}
\label{eq:K_example}
K = \left[\begin{array}{cc;{2pt/2pt}cccc;{2pt/2pt}ccc|c}
  \text{-}k_1^t  &   0  & k_1^t  &   0   &  0  &   0 &    0   &  0   &  0& 0\\
     0  &  \text{-}k_1^r  &   0   &  k_1^r  &   0  &   0   &  0   &  0  &   0 &0\\
     \hdashline[2pt/2pt]
   k_1^t    & 0 & \text{-}k_1^t   &  0   &  0  &   0  &   0   &  0 &    0 &0\\
     0   &  k_1^r   &  0  &  \text{-}k_1^r  &   0  &   0 &    0 &    0   &  0 &0\\
     0   &  0  &   0  &   0   &  0  &   0  &   0  &   0  &  0   & 1\\
     0   &  0  &   0  &   0 & \text{-}k_2^t  &   0  & k_2^t  &  0   &  0 &0\\
     0   &  0  &   0  &   0 &    0  &  \text{-}k_2^r &    0  &   k_2^r  &   0 &0\\
     \hdashline[2pt/2pt]
     0    & 0  &   0  &   0 &  k_2^t &   0 & \text{-}k_2^t  &   0   &  0 &0\\
     0   &  0  &   0  &   0  &   0    & k_2^r  &   0 &   \text{-}k_2^r   &  0 &0 \\
\hline 
     0  &   0   &  0   &  0   &  0  &   0  &   0  &  0  &   1  &  0 \\
\end{array} \right].
\end{equation}
The external input force $u_c$ [N] is applied to the middle of subsystem 2 in the transversal direction.
The external output displacement $y_c$ [m] is measured at the middle of subsystem 3 in the transversal direction.

Each beam/subsystem is discretized by linear two-node Euler beam elements (only bending, no shear, see~\cite{craig2006}) of equal length. 
Per node we have one translational degree of freedom (dof), i.e., a transversal displacement, and one rotational dof).
For each beam, viscous damping is modelled using 6\% modal damping.
Physical and geometrical parameter values of the three beams and information about finite element discretization, the number of states and the number of subsystem inputs and outputs are given in Table~\ref{tab:parameters}.
With this information we can construct $G_1(s)$, $G_2(s)$ and $G_3(s)$ and the interconnected system $G_c(s)$ is then given according to Section~\ref{sec:Problem} where $k=3$ and $K$ is defined by (\ref{eq:K_example}).
\begin{table}[]
\caption{Parameter values of each subsystem in the example system. 
In addition, information about finite element discretization, the state-space dimensions of a minimal realization of the high-order subsystems, and the number of inputs and outputs per subsystem are specified.}
\label{tab:parameters}
\begin{tabular}{l|lll}
Parameter                             & Subsys. 1       & Subsys. 2       & Subsys. 3       \\
\hline                        
Cross-sect. area [m$^2$]       & $1\times 10^{\text{-}5}$ & $1\times 10^{\text{-}5}$ & $1\times 10^{\text{-}5}$ \\
2nd area moment [m$^4$]        & $1\times 10^{\text{-}9}$ & $1\times 10^{\text{-}9}$ & $1\times 10^{\text{-}9}$ \\
Young's modulus [Pa]           & $2\times 10^{11}$        & $2\times 10^{11}$        & $2\times 10^{11}$ \\
Mass density [kg/m$^3$]        & $8\times 10^{3}$         & $8\times 10^{3}$         & $8\times 10^{3}$  \\
Modal damping [-]              & $0.06$                   & $0.06$                   & $0.06$  \\
Length [m]                     & $1$                      & $0.4$                    & $0.6$             \\
\# of elements [-]             & $100$                    & $40$                     & $60$              \\
\hline
Transfer function              & $G_1(s)$                 & $G_2(s)$                 & $G_3(s)$             \\             
\hline                              
\# of states $n_j$ [-]         & $400$             	      & $164$                    & $240$       \\   
\# of inputs $m_j$ [-]         & $2$                      & $5$                      & $2$       \\    
\# of outputs $p_j$ [-]        & $2$                      & $4$                      & $3$            
\end{tabular}
\end{table}

Below, both the bottom-up and the top-down approaches from Section~\ref{sec:Error} are concisely illustrated.
First, in Section~\ref{sec:Bottom}, we consider a bottom-up problem in which we show how the error in the interconnected system introduced by the reduction of subsystem $1$ can be bounded first by using a global error bound and then by using a frequency-dependent error bound.
Subsequently, in Section~\ref{sec:Top}, we show how specifications on the accuracy of the reduced-order interconnected system can be translated to frequency-dependent bounds on the reduction error of subsystem $1$ and how this information can be used to find a ROM that takes this frequency-dependent bound into account.

\subsection{Bottom-up approach}
\label{sec:Bottom}
In the bottom-up approach, we use Theorems~\ref{the:BU_GLOBAL} and \ref{the:BU_FREQ} to find both global and frequency-dependent a priori bounds on the reduction error of the interconnected system model shown in Figure~\ref{fig:example}.
Note that in this example, subsystem $1$ is reduced using balanced truncation~\cite{antoulas2005} to find some $\hat{G}_1(s)$ whereas the models of subsystems $2$ and $3$ are left unreduced.
\begin{table*}[]
\centering
\caption{Bottom-up global error bounds example: comparison between $\|E_c\|_\infty \leq \bar{\varepsilon}_{c,a} \leq \bar{\varepsilon}_c$ for a different number of states $r_1$ in the reduced-order subsystem $\hat{G}_1$.
To show conservativeness of the error bounds, $\bar{\varepsilon}_{c,a}/\|E_c\|_\infty$ and $\bar{\varepsilon}_{c}/\|E_c\|_\infty$ are given.
$\bar{\varepsilon}_{1}/\|E_1\|_\infty$ shows the conservativeness present in the a priori error bound on subsystem level.
If no feasible solution for (\ref{eq:sdp_bottom_up}) could be found, the result is denoted with "-".}
\label{tab:BU}
\begin{tabular}{l|l|ll|ll|l}
$r_1 $ [-]& $\|E_c\|_\infty$ [m/N] & $\bar{\varepsilon}_{c,a}$ [m/N] & $\bar{\varepsilon}_c$ [m/N] & $\bar{\varepsilon}_{c,a}/\|E_c\|_\infty$ [-] & $\bar{\varepsilon}_{c}/\|E_c\|_\infty$ [-] & $\bar{\varepsilon}_{1}/\|E_1\|_\infty$ [-]\\
\hline
500 ($n_1$)   & $0$                    & $0$                   & $0$  				  & - 					  & -					 & - \\
\hline
140   & $2.59\times 10^{-9}$   & $3.52\times 10^{-9}$  & $1.03\times 10^{-7}$ & $1.36$   & $40.0$  & $2.25$  \\
120   & $5.01\times 10^{-9}$   & $6.02\times 10^{-9}$  & $5.74\times 10^{-7}$ & $1.20$   & $115$   & $6.46$  \\
100   & $1.89\times 10^{-8}$   & $2.90\times 10^{-8}$  & $3.43\times 10^{-6}$ & $1.53$   & $181$   & $99.2$  \\
80    & $4.32\times 10^{-8}$   & $1.18\times 10^{-7}$  & $3.46\times 10^{-5}$ & $2.73$   & $801$   & $163$  \\
60    & $9.04\times 10^{-8}$   & $5.44\times 10^{-7}$  & -           		  & $6.01$   & -       & $223$  \\
40    & $2.19\times 10^{-6}$   & $2.30\times 10^{-5}$  & -           		  & $10.5$   & -       & $56.0$  \\
20    & $1.00\times 10^{-4}$   & -         		       & -           		  & -   	 & -       & $9.25$  \\
\end{tabular}
\end{table*}
First, we consider two types of errors on $E_1$:
\begin{enumerate}
\item A priori error bounds $\|E_1\|_\infty \leq \bar{\varepsilon}_1$ using the Hankel singular values as in (\ref{eq:apriori}). 
\item The actual $\mathcal{H}_\infty$-norm of the error dynamics $\|E_1\|_\infty = \bar{\varepsilon}_{1,a}$, determined after the reduction of the subsystem.
\end{enumerate}
\begin{rem}
Note that $\bar{\varepsilon}_1$ is used to determine a priori error bounds $\bar{\varepsilon}_c$ on the interconnected system.
However, these a priori error bounds already have some conservativeness on a subsystem level.
By using the actual $\mathcal{H}_\infty$-norm of the error dynamics $\bar{\varepsilon}_{1,a}$, the conservativeness that is a result of Theorem~\ref{the:BU_GLOBAL} can be determined.
\end{rem}
These errors are computed for varying values of the reduced-order $r_1$ of subsystem 1. 
Second, the bottom-up SDP problem in Theorem~\ref{the:BU_GLOBAL} is solved for the resulting values for $\bar{\varepsilon}_1$ and $\bar{\varepsilon}_{1,a}$.
The solution to these SDPs provides an error bound $\|E_c\|_\infty \leq \bar{\varepsilon}_c$ and $\|E_c\|_\infty \leq \bar{\varepsilon}_{c,a}$ for $\bar{\varepsilon}_1$ and $\bar{\varepsilon}_{1,a}$, respectively.
Finally, to compare how conservative the error bounds are with respect to different orders of reduction, $\bar{\varepsilon}_{c,a}/\|E_c\|_\infty$ and $\bar{\varepsilon}_{c}/\|E_c\|_\infty$ are determined.
The results are given in Table~\ref{tab:BU}.
\begin{rem}
Note that the actual $\|E_c\|_\infty$ can only be computed a posteriori (so after the reduction has been pursued and the reduced-order interconnected system has been constructed). We emphasize that the methodology in this paper allows to compute the bounds for $\|E_c\|_\infty$ a priori.
\end{rem}
\begin{figure}
   \centering
   \includegraphics[scale=.9]{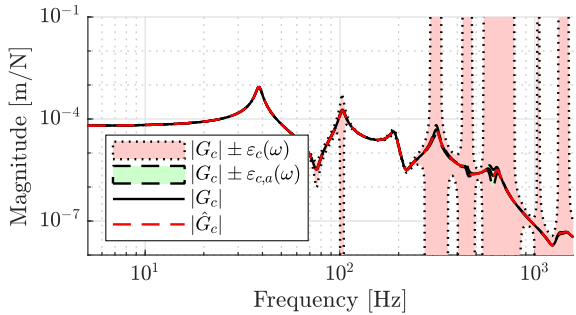} 
   \caption{Bottom-up approach: Magnitude plot for the high-order interconnected system $|G_c(i\omega)|$, the reduced-order interconnected system $|\hat{G}_c(i\omega)|$ with $r_1 = 60$, the a priori error bound using a priori frequency-dependent subsystem errors $|G_c(i\omega)| \pm \varepsilon_c(\omega)$ and the frequency-dependent error bound using the largest singular values of the actual subsystem error $|G_c(i\omega)| \pm \varepsilon_{c,a}(\omega)$.}
   \label{fig:BU_Gc}
\end{figure}
\begin{figure}
   \centering
   \includegraphics[scale=.9]{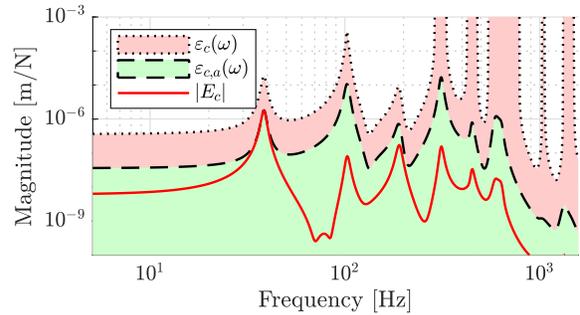} 
   \caption{Bottom-up approach: Magnitude plot for the interconnected system error dynamics $|E_c(i\omega)|$ with $r_1 = 60$, the a priori frequency-dependent error bound using a priori subsystem errors $\varepsilon_c(\omega)$ and the frequency-dependent error bound using the largest singular values of the actual subsystem error $\varepsilon_{c,a}(\omega)$.}
   \label{fig:BU_Ec}
\end{figure}
From Table~\ref{tab:BU}, we can make several observations.
Solving the bottom-up SDP problem in Theorem~\ref{the:BU_GLOBAL} provides an a priori global error bound on the error dynamics of the interconnected system $\|E_c\|_\infty$ using global error bounds on subsystem level.
Additionally, for the $\mathcal{H}_\infty$-norm of the actual error dynamics $\bar{\varepsilon}_{c,a}$, the bottom-up approach provides tight(er) error bounds on the level of the interconnected system, as indicated by the small values of $\bar{\varepsilon}_{c,a}/\|E_c\|_\infty$.
In contrast, since the values of $\bar{\varepsilon}_{c}/\|E_c\|_\infty$ are clearly higher, the error bound using the a priori errors on the subsystem level $\|E_1\|_\infty \leq \bar{\varepsilon}_{1}$ is significantly more conservative.
However, the subsystem a priori error bound $\bar{\varepsilon}_{1}$ already provides some level of conservativeness, as indicated by the values of $\bar{\varepsilon}_{1}/\|E_1\|_\infty$. 
Therefore, we postulate that the conservativeness of the a priori error bound $\bar{\varepsilon}_{c}/\|E_c\|_\infty$ is for a significant part attributed to the conservativeness of $\bar{\varepsilon}_{1}/\|E_1\|_\infty$.

As can be seen in Table~\ref{tab:BU}, for this example, no global a priori error bound $\bar{\varepsilon}_c$ can be found for a reduction of $r_1 \leq 60$ using Theorem~\ref{the:BU_GLOBAL} with $\bar{\varepsilon}_{1}$.
However, although no global error bound on $E_c$ can be found for significant reduction of subsystem 1, frequency-dependent bounds can still be found for intervals $\omega \in \R$.
Below, we show how Theorem~\ref{the:BU_FREQ} can be used to find frequency-dependent error bounds on $E_c$.
Frequency-dependent error bounds can be computed to provide useful insights on how reduction errors on the subsystem level propagate to the interconnected system.
\begin{figure}
   \centering
   \includegraphics[scale=.9]{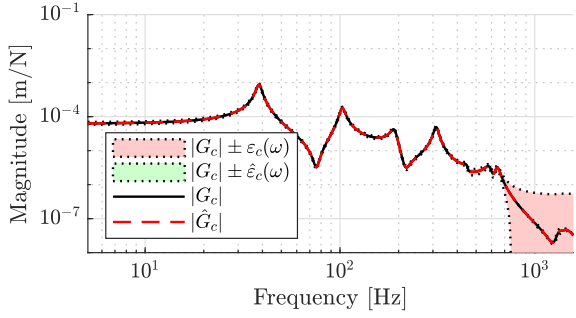} 
   \caption{Top-down approach: Magnitude plot for the high-order interconnected system $|G_c(i\omega)|$, the reduced-order interconnected system $|\hat{G}_c(i\omega)|$ with $r_1 = 20$, the a priori frequency-dependent error bound using the actual subsystem error bounds $|G_c(i\omega)| \pm \hat{\varepsilon}_c(\omega)$ and the user defined frequency-dependent error bound $|G_c(i\omega)| \pm \varepsilon_{c}(\omega)$.}
   \label{fig:TD_Gc}
\end{figure}
\begin{figure}
   \centering
   \includegraphics[scale=.9]{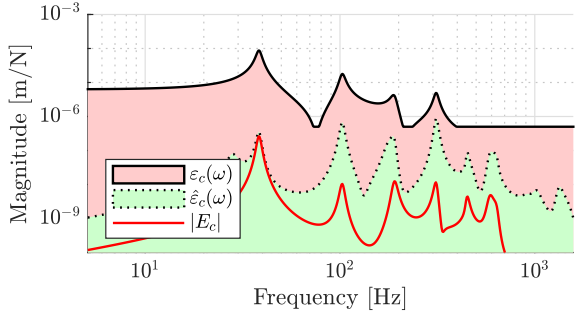} 
   \caption{Top-down approach: Magnitude plot for the high-order interconnected error dynamics $|E_c(i\omega)|$ with $r_1 = 20$, the frequency-dependent error bound using the actual subsystem errors $\hat{\varepsilon}_c(\omega)$ and the user defined frequency-dependent error bound $\varepsilon_{c}(\omega)$.}
   \label{fig:TD_Ec}
\end{figure}

First, we compute a reduced-order model $\hat{G}_1$ for subsystem 1 using balanced truncation with $r_1 = 60$.
Then, we consider and calculate two types of errors on $E_1(s)$:
\begin{enumerate}
\item The same a priori error bounds $\|E_1\|_\infty \leq \bar{\varepsilon}_1$ as in the global case (see Remark~\ref{rem:bt_fd}).
\item Frequency-dependent error bounds $\varepsilon_{1,a}(\omega)$ as the largest singular value of the actual error dynamics $\bar{\sigma}(E_1(i\omega)) = \varepsilon_{1,a}(\omega)$, determined after the reduction of the subsystem.
\end{enumerate}
Second, the bottom-up SDP problem in Theorem~\ref{the:BU_FREQ} is solved for the resulting values for $\varepsilon_{1}(\omega) = \bar{\varepsilon}_{1}$ and $\varepsilon_{1,a}(\omega)$.
The solution to these SDPs provides an error bound $\bar{\sigma}(E_c(i\omega)) \leq \varepsilon_c(\omega)$ and $\bar{\sigma}(E_c(i\omega)) \leq \varepsilon_{c,a}(\omega)$ for each $\varepsilon_{1}(\omega) =\bar{\varepsilon}_{1}$ and $\varepsilon_{1,a}(\omega)$, respectively, as shown in Figures~\ref{fig:BU_Gc} and~\ref{fig:BU_Ec}.
\begin{figure}
   \centering
   \includegraphics[scale=.9]{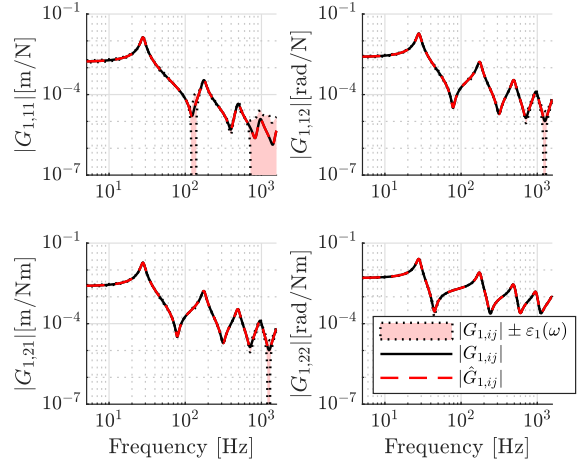} 
   \caption{Top-down approach: Magnitude plot of the high-order subsystem $|G_{1,ij}(i\omega)|$, the reduced-order subsystem $|\hat{G}_{1,ij}(i\omega)|$ with $r_1 = 20$ and the allowed frequency-dependent error with the top-down approach $|G_{1,ij}(i\omega)| \pm \varepsilon_1(\omega)$ from input $i$ to output $j$.}
   \label{fig:TD_G1}
\end{figure}
\begin{figure}
   \centering
   \includegraphics[scale=.9]{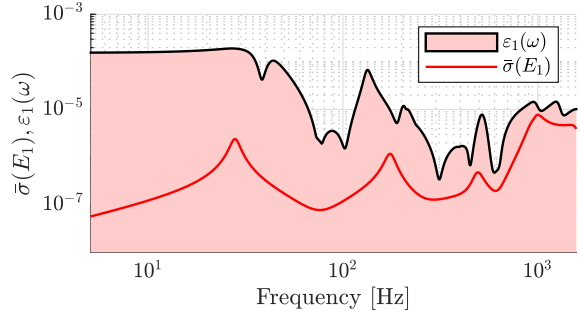} 
   \caption{Top-down approach: Largest singular value plot of the high-order subsystem error dynamics $\bar{\sigma}(E_1(i\omega))$ with $r_1 = 20$ and the allowed frequency-dependent error found using the top-down approach $\varepsilon_1(\omega)$.}
   \label{fig:TD_E1}
\end{figure}

From these figures, we can make several observations.
In Figures~\ref{fig:BU_Gc} and~\ref{fig:BU_Ec}, in the green areas, the frequency-dependent a priori error bound $\varepsilon_{c,a}(\omega)$ (based on $\varepsilon_{1,a}(\omega)$) is found for all frequencies in the shown domain.
In Figure~\ref{fig:BU_Ec}, it can be clearly seen that this bound is particularly tight.
Therefore, in Figure~\ref{fig:BU_Gc}, the effect of $\varepsilon_{c,a}(\omega)$ is not even visible.
Additionally, it can be seen in the red areas, which illustrate the frequency-dependent a priori error bound bound $\varepsilon_c(\omega)$, that for several frequencies in the shown domain, no bound $\varepsilon_c(\omega)$ is found.
This is in agreement with Table~\ref{tab:BU}, where the bottom-up approach indeed cannot find a global a priori error bound $\varepsilon_c(\omega)$ for $r_1 = 60$. 
However, for most other frequencies, an a priori error bound $\varepsilon_c(\omega)$ can still be computed.
In total, both $\varepsilon_{c}$ and $\varepsilon_{c,a}$ give a clear frequency-dependent reduction error bound for the largest part of the frequency domain, which can be used to give a frequency-dependent upper bound on how substructure reduction errors propagate to the interconnected system, even when no global error bound can be found.
The inverse of this problem, finding an upper bound on subsystem errors based on requirements on the interconnected system, is shown on the same system in the next section.

\subsection{Top-down approach}
\label{sec:Top}
In this example, the top-down approach using the optimization problem in Theorem~\ref{the:TD_FREQ} is applied to the example system in Figure~\ref{fig:example}.
Specifically, a required frequency-dependent reduction error bound $\bar{\sigma}(E_1(i\omega)) \leq \varepsilon_1(\omega)$ for subsystem 1 is computed that guarantees a user-selected frequency-dependent error bound $\bar{\sigma}(E_c(i\omega)) \leq \varepsilon_c(\omega)$ on the interconnected system.
Additionally, a reduced-order subsystem $\hat{G}_1(s)$ is computed that meets these requirements.
Recall that subsystems 2 and 3 remain unreduced, i.e., $\bar{\varepsilon}_2 = \varepsilon_2(\omega) = \bar{\varepsilon}_3 = \varepsilon_3(\omega) =0$. 
The computation of $\hat{G}_1(s)$ serves two purposes.
Namely, 1) to validate the results, and 2) to show how frequency-weighted balanced truncation can exploit the frequency information in the error bounds to increase the amount of reduction that can be achieved.
\begin{rem}
\label{rem:FWBT}
Note that any reduced-order subsystem $\hat{G}_1(s)$ for which $\bar{\sigma}(E_1(i\omega)) \leq \varepsilon_1(\omega)$ holds for the specified frequencies can be used, regardless of the model reduction method.
Frequency-weighted balanced truncation is a special form of balanced truncation in which the goal is to reduce the frequency-weighted error between $G(s)$ and $\hat{G}(s)$~\cite{gugercin2004}.
This method is particularly suitable for the top-down approach since it allows to capitalize on the computed frequency-dependent error bound $\varepsilon_1(\omega)$ for the reduction of the subsystem.
Specifically, we can directly apply the computed $W^F$ in Theorem~\ref{the:TD_FREQ} as a weighting for the reduction.
In this example, we use Enns' method~\cite{enns1984} to minimize $\|W^F_1 E_1\|_\infty$ (whereas regular balanced truncation minimizes $\|E_1\|_\infty$).
\end{rem}
The top-down approach is applied in this example by carrying out the following steps:
\begin{enumerate}
\item All frequencies $\omega$ over a grid of 1000 logarithmically spaced points in the interval $[10^{1.5},10^4]$ rad/s are evaluated.
For these frequencies, a frequency-dependent error bound $\varepsilon_c(\omega)$ is defined, as can be done by the user (see Remark~\ref{rem:e_select}). 
In this example, this bound is chosen as some fraction $\beta_1$ of the magnitude of $\bar{\sigma}(G_c(i\omega))$, bounded below by $\beta_2$, given as 
\begin{equation}
\varepsilon_c(\omega) = \max\{\beta_1 \cdot \bar{\sigma}(G_c(i\omega)), \beta_2\},
\end{equation} 
where $\beta_1=0.1$ and $\beta_2=5\times 10^{-7}$ in this example.
In Figures~\ref{fig:TD_Gc} and \ref{fig:TD_Ec}, this bound $\varepsilon_c(\omega)$ is indicated by the red areas.
\item The SDP problem in Theorem~\ref{the:TD_FREQ} is solved using $\varepsilon_c(\omega)$ with $q=1$ to find some frequency-dependent error bound $\varepsilon_1(\omega)$ on the error dynamics of the first subsystem.
Note that we assume unreduced subsystems 2 and 3, and therefore $\varepsilon_2(\omega) = \varepsilon_3(\omega) = 0$.
The error bound $\varepsilon_1(\omega)$ corresponding to $\varepsilon_c(\omega)$ is indicated by the red areas in Figures~\ref{fig:TD_G1} and \ref{fig:TD_E1}.
\item A reduced-order subsystem $\hat{G}_1$ with $r_1=20$ is found using frequency-weighted balanced truncation (see Remark~\ref{rem:FWBT}) for which $\bar{\sigma}(E_1(i\omega)) \leq \varepsilon_1(\omega)$ holds, as shown by the red lines in Figures~\ref{fig:TD_G1} and \ref{fig:TD_E1}.
\item For validation, the bottom-up approach as given in Theorem~\ref{the:BU_FREQ} is solved to find some a priori worst-case upper bound on the interconnected system error dynamics $\bar{\sigma}(E_c(i\omega)) \leq \hat{\varepsilon}_c(\omega)$ caused by the replacement of $G_1$ by $\hat{G}_1$ in the interconnected system.
In Figures~\ref{fig:TD_Gc} and \ref{fig:TD_Ec}, this bound $\hat{\varepsilon}_c(\omega)$ is indicated by the green areas.
\item As additional validation, the reduced-order interconnected system $\hat{G}_c$ and the error dynamics $E_c$ are computed and shown by the red line in Figures~\ref{fig:TD_Gc} and \ref{fig:TD_Ec}, which are indeed fully within the red and green areas.
\end{enumerate}
\begin{rem}
Note that in theory, if $\bar{\sigma}(E_1(i\omega))$ would match the allowed error bound $\varepsilon_1(\omega)$ in Figure~\ref{fig:TD_E1}, we would see that $\hat{\varepsilon}_c(\omega) = \varepsilon_{c}(\omega)$ in Figure~\ref{fig:TD_Ec}.
However, after reduction, Figure~\ref{fig:TD_E1} shows that the largest singular values $\bar{\sigma}(E_1(i\omega))$ of the reduced-order subsystem 1 do not fully ``utilize" the allowed error $\varepsilon_1(\omega)$.
As a result, in Figure~\ref{fig:TD_Ec}, we can see that $\hat{\varepsilon}_c(\omega)$ is much smaller than $\varepsilon_{c}(\omega)$.
Therefore, the distance between $|E_c(i\omega)|$ and $\varepsilon_{c}(\omega)$ in Figure~\ref{fig:TD_Ec} is a combination between the conservativeness of the top-down approach, given by the gap in $|E_c(i\omega)| < \hat{\varepsilon}_c(\omega)$, and the fact that $\bar{\sigma}(E_1(i\omega)) \leq \varepsilon_{1}(\omega)$, which results in the gap in $\hat{\varepsilon}_c(\omega) \leq \varepsilon_{c}(\omega)$.
\end{rem}
In summary, this example shows that the top-down approach can be effectively used to find subsystem 1 error bounds given some interconnected system accuracy specification.
This translation from requirements on the interconnected system to requirements on a subsystem level is particularly useful because 1) any reduced-order subsystem model that satisfies the bounds is guaranteed not to cause the error in the interconnected system to exceed the required accuracy, and 2) a reduced-order subsystem model can be developed by making use of the frequency-dependent error bound, in this case using frequency-weighted balanced truncation (see Remark~\ref{rem:FWBT}), to further reduce the interconnected system model.

\section{Conclusions}
\label{sec:Conclusion}
Modular model reduction is a computationally efficient method that allows for the computation of ROMs of interconnected (multidisciplinary and multi-physical) subsystems.
However, generally, modular model reduction leads to less accurate ROMs of the interconnected system in comparison to costly direct (structure-preserving) reduction methods.
In this paper, to mitigate this accuracy disadvantage, a mathematical relation between the accuracy of reduced subsystem models and the accuracy of the reduced interconnected system model is introduced.

The main idea relies on defining the error dynamics introduced by the MOR of a subsystem as a block-diagonal structured uncertainty.
Then, the system can be reformulated into the framework of a robust performance problem.
This allows for a direct computation of a relation between upper bounds of subsystem reduction error dynamics to upper bounds on the interconnected system reduction error dynamics using the structured singular value $\mu$.

This relation can then be used in the two ways.
1) a bottom-up approach can be used to guarantee stability of the interconnected, reduced-order system and determine (frequency-dependent) a priori error bounds for interconnected system model reduction when a priori error bounds are available for the reduced subsystem models.
2) a top-down approach allows the user to define (frequency-dependent) accuracy specifications on the reduced interconnected system model.
These specifications can then be translated to (frequency-dependent) accuracy requirements on reduced subsystem models. 
When these are achieved, they guarantee that the user-defined specifications on the interconnected system hold.
Additionally, this allows for the effective use of frequency-weighted balanced truncation to achieve reduction of the subsystem while guaranteeing that the interconnected system accuracy specifications are met and remains stable.
To demonstrate the use of these approaches, they have been applied to a structural dynamics beam system.


\ack
This publication is part of the project Digital Twin with project number P18-03 of the research programme Perspectief which is (mainly) financed by the Dutch Research Council (NWO).

\bibliographystyle{unsrt}        
\bibliography{refs}              

\end{document}